\documentstyle[preprint,aps]{revtex}
\begin{document}
\draft
\preprint{CLNS 95/1371}
\title{Charmed Strange Pentaquarks in the Large $N_c$ Limit}
\author{Chi-Keung Chow}
\address{Newman Laboratory of Nuclear Studies, Cornell University, Ithaca,
NY 14853.}
\date{\today}
\maketitle
\begin{abstract}
The properties of pentaquarks containing a heavy anti-quark and strange
quarks are studied in the bound state picture.
In the flavor SU(3) limit, there are many pentaquark states with the same
binding energy.
When the SU(3) symmetry breaking effects are included, however, three states
become particularly stable due to a ``Gell-Mann--Okubo mechanism''.
They are the $\bar Qsuud$ and $\bar Qsudd$ states discussed by Lipkin, and
a previously unstudied $\bar Qssud$ state.
These states will have $J^P={1\over2}^+$ and their masses are estimated.
These states, if exist, may be seen in experiments in the near future.
\end{abstract}
\narrowtext
Quark confinement mandates that hadrons must be singlets under SU(3) color.
Ordinary hadrons, i.e., mesons as $q\bar q$ bound states and baryons as
$qqq$ bound states, do conform to this rule.
On the other hand, it is also possible to construct other SU(3) singlet
quark states.
The most famous of these multiquark exotic states are the Jaffe tetraquark
$qq\bar q\bar q$ \cite{1,2} and hexaquark $qqqqqq$ \cite{3}.
Five quark bound states $\bar qqqqq$, known as pentaquarks, are first
suggested in Ref.~\cite{4} and are subsequently studied in
Ref.~\cite{5,6,7,8,9,10}.
In particular, Lipkin \cite{5,9} suggested that the states $\bar Qsuud$
and $\bar Qsudd$, where $Q$ is a heavy quark ($c$ or $b$), are especially
stable under the ``flavor antisymmetry principle''.
These states, if exist, may be discovered soon in the Fermilab E791
experiment.

In a previous article \cite{10}, the author has discussed the possibility
of constructing heavy pentaquarks $\bar Qqqqq$ in the large $N_c$ limit.
These pentaquarks appear as bound states of heavy mesons to chiral
{\it antisolitons}.
However, that investigation was made under SU(2)$_L\;\times\;$SU(2)$_R$
chiral symmetry, in which all the states have zero strangeness.
In order to take the strange quark into account, we have to incorporate
SU(3) flavor symmetry into our model, which is the objective of this
article.
We will see that, when SU(3) flavor symmetry is unbroken, there will be
a large number of degenerate pentaquark states, all of them weakly bounded
but may be destabilized by higher order $1/M_Q$ and/or $1/N_c$ corrections.
When, however, the SU(3) flavor symmetry breaking effect is taken into
account, the states studied by Lipkin will be even more tightly bounded by
a ``Gell-Mann--Okubo mechanism'' analogous to which cause the $\Sigma-
\Lambda$ splitting.
As a result, these states will have the best chances to survive the higher
order corrections and remain bounded in the real world.

It will help to review the relevant results in Ref. \cite{10}.
In the large $N_c$ limit, the nucleon N and the Delta $\Delta$ appear as
topological solitons of the SU(2) pion fields, which are the Goldstone
field of the spontaneous symmetry breaking SU(2)$_L\;\times\;$SU(2)$_R\to$
SU(2)$_V$ \cite{11}.
Then the interaction of such chiral solitons with heavy mesons can be
studied under the chiral lagrangian \cite{12,13,14}.
It turns out that only states with $K=I+s_\ell=0$ are bound \cite{15,16,17}.
The binding energy $V$ is
\begin{equation}
V_0=V(K=0)=-\textstyle{3\over2}gF'(0),
\label{scale}
\end{equation}
where $g$ is the axial current coupling constant in the chiral lagrangian
and $F'(0)$ is the derivative of the soliton profile function at the
origin.
Since $F'(0)>0$, the negative sign in Eq. (\ref{scale}) implies that the
binding energy is negative and the resultant state is really bounded.
These states can be identified as $\Lambda_Q$ with $I=s_\ell=0$ and
$\Sigma_Q$ with $I=s_\ell=1$.
On the other hand, the $K=1$ states are unbounded, with binding energy
\begin{equation}
V(K=1)=-\textstyle{1\over3}V_0.
\end{equation}

On the other hand, pentaquarks merge naturally as heavy meson--chiral
antisoliton bound states in the same picture.
If only terms with at most one derivative are retained in the chiral
lagrangian (the truncated lagrangian in Ref.~\cite{10}), the binding energy,
denoted by $\tilde V$ in this case, is just $V$ with the opposite sign.
\begin{equation}
\tilde V(K)=-V(K).
\end{equation}
Hence, the pentaquark bound states are those with $K=1$, i.e., $(I,s_\ell)=
(1,0)$, $(1,1)$ or $(0,1)$, with binding energy a third that of a normal
heavy baryon.
\begin{equation}
\tilde V(K=1)=\textstyle{1\over3}V(K=0)=\textstyle{1\over3}V_0.
\end{equation}

Under flavor SU(3), the nucleon isodoublet becomes a part of the ground
state baryon octet.
We will follow the notation of section 7 of Ref.~\cite{15} and denote an
irreducible SU(3) representation as $(m,n)$, which is a traceless tensor
completely symmetric in $m$ upper and $n$ lower indices, and have
dimension
\begin{equation}
\dim(m,n)=\textstyle{1\over2} (m+1)(n+1)(m+n+2).
\end{equation}
For example, the fundamental triplet {\bf 3} is $(1,0)$, sextet {\bf 6} is
$(2,0)$, octet {\bf 8} (1,1), decuplet {\bf 10} (3,0), {\it etc.}
The conjugate representations are obtained by interchanging $m$ and $n$,
i.e., antitriplet {\bf $\bar 3$} is $(0,1)$, {\it etc.}
Note that there are four irreducible representations of dimension 15:
$(4,0)$, $(2,1)$ and their conjugates.

Now we are ready to generalize the SU(2) results above to SU(3) \cite{15}.
First we will generalize the results for the normal heavy baryons.
\begin{mathletters}
The baryon octet--heavy meson antitriplet bound states are,
\begin{eqnarray}
(1,1)\otimes(0,1)&=&(0,1)\oplus(2,0)\oplus(1,2)\nonumber\\
{\bf 8}\otimes{\bf \bar 3}&=&{\bf \bar 3}\oplus{\bf 6}\oplus{\bf \bar {15}},
\label{b8}
\end{eqnarray}
while the baryon decuplet--heavy meson antitriplet bound states are,
\begin{eqnarray}
(3,0)\otimes(0,1)&=&(2,0)\oplus(3,1)\nonumber\\
{\bf 10}\otimes{\bf \bar 3}&=&{\bf 6}\oplus{\bf 24}.
\label{b10}
\end{eqnarray}
\label{b}
\end{mathletters}
The stable states are those connected to the states stable in SU(2) by an
SU(3) rotation.
For example, the $I=s_\ell=0$ $\Lambda_Q$ state is an element of the
$(0,1)$ in Eq.~(\ref{b8}) and hence the whole antitriplet, with $s_\ell=0$,
will have binding energy $V_0$.
Physically it represents the $(\Lambda_Q, \Xi_Q)$ antitriplet.
Similarly, $\Sigma^{(*)}_Q$, with $I=s_\ell=1$, is an element of a
particular combination of the $(2,0)$ in Eq.~(\ref{b8}) (denoted as
$(2,0)_8$) and that in Eq.~(\ref{b10}) (denoted as $(2,0)_{10}$).
\begin{equation}
\Sigma_Q\in (2,0)_\Sigma=\sqrt{1\over3}(2,0)_8+\sqrt{2\over3}(2,0)_{10}.
\end{equation}
Thus the whole $(2,0)_\Sigma$ sextet, with $s_\ell=1$ will have binding
energy $V_0$.
It corresponds to the $(\Sigma^{(*)}_Q, \Xi^{'(*)}_Q, \Omega^{(*)}_Q)$
sextet in the real world.

A similar analysis can be done in the pentaquarks sector.
\begin{mathletters}
Since the heavy anti-mesons form a triplet under flavor SU(3), the
counterparts of Eqs.~(\ref{b}) are,
\begin{eqnarray}
(1,1)\otimes(1,0)&=&(1,0)\oplus(0,2)\oplus(2,1)\nonumber\\
{\bf 8}\otimes{\bf 3}&=&{\bf 3}\oplus{\bf \bar 6}\oplus{\bf 15},
\label{p8}
\end{eqnarray}
\begin{eqnarray}
(3,0)\otimes(1,0)&=&(2,1)\oplus(4,0)\nonumber\\
{\bf 10}\otimes{\bf 3}&=&{\bf 15}\oplus{\bf 15'}.
\label{p10}
\end{eqnarray}
Note that the two ${\bf 15}$'s in Eq.~(\ref{p10}) are neither equivalent nor
conjugate to each other.
\end{mathletters}
Recall that, in the SU(2) case, we have identified three stable bound states
with $(I,s_\ell)=(0,1)$, $(1,1)$ and $(1,0)$.
The $(I,s_\ell)=(0,1)$ state now is a part of the $(0,2)$ representation in
Eq.~(\ref{p8}), while $(I,s_\ell)=(1,0)$ state is inside the $(2,1)$ in
Eq.~(\ref{p8}).
The $(I,s_\ell)=(1,1)$ state falls in the linear combination of the $(2,1)$'s
in Eq.~(\ref{p8}) and that in Eq.~(\ref{p10}) shown below.
\begin{equation}
(2,1)_\Sigma=\sqrt{2\over3}(2,1)_8-\sqrt{1\over3}(2,1)_{10}.
\end{equation}
In the large $N_c$ limit, all these states will be degenerate.
In the real world, the baryon decuplet is heavier than the octet by
$M_\Delta-M_N\sim 300$ MeV.
This would cause the $(2,1)_\Sigma$ state to be heavier than the other two
states by ${1\over3}(M_\Delta-M_N)\sim 100$ MeV.
(This is the same mechanism which breaks the $\Sigma^{(*)}_Q-\Lambda_Q$
degeneracy,
$M_{\Sigma^{(*)}_Q}-M_{\Lambda_Q}={2\over3}(M_\Delta-M_N)$.)
Since pentaquarks are only weakly bounded (if bounded at all) in all existing
models, it is probable that this 100 MeV mass difference will destabilize the
$(2,1)_\Sigma$ pentaquarks.
Hence we will ignore this state for the rest of our discussion.
This leaves us with the $s_\ell=1$ ${\bf \bar 6}$ and the $s_\ell=0$
{\bf 15}, both just involving the baryon octet and hence degenerate (up to
order $N_c^0$) even at finite $N_c$.
All of them will have binding energy $-{1\over3}V_0$.
We will see that the flavor SU(3) symmetry breaking is going to pick out the
Lipkin states (and one other state) as the most tightly bounded ones.

To investigate the effect of SU(3) symmetry breaking effects, it helps to
review the SU(3) symmetry breaking in the baryon octet, in which the baryon
masses are governed by the Gell-Mann--Okubo formula \cite{18,19},
\begin{equation}
M=M_0+aY+b(I(I+1)-\textstyle{1\over4}Y^2).
\end{equation}
The last term is responsible for the physical $\Sigma-\Lambda$ splitting,
without which all four $S=1$ baryons will be degenerate and be exactly
half way between N and $\Xi$ on the mass spectrum.
In the real world, $b$ is positive and hence $\Sigma$ is heavier than
$\Lambda$ by about 78 MeV.
Since the Gell-Mann--Okubo formula is the consequence of just SU(3) group
theory, a similar formula with different coefficients also governs the
pentaquark masses in each of the representations.
Then this ``Gell-Mann--Okubo mechanism'' may also break the degeneracy
between states with the same strangeness in the $(2,1)$  representation.
In particular, for $S=1$, there exists both an isodoublet and an isoquartet
in the $(2,1)$ representation.
The octet example leads us to expect the isodoublet to be lowered and the
isoquartet raised.
If this is indeed the case, the most stable bound states will be the
$\bar Qsuud$ and $\bar Qsudd$ isodoublet, i.e., exactly the states Lipkin
predicted.

To verify this conjecture, one must project the Lipkin states back into
the heavy meson--octet baryon product space.
In general,
\begin{equation}
|\bar Qsuud\rangle_L = w|\bar Qs\rangle |p\rangle
+ x|\bar Qd\rangle |\Sigma^+\rangle + y |\bar Qu\rangle |\Sigma^0\rangle
+ z|\bar Qu\rangle |\Lambda\rangle,
\end{equation}
with the normalization condition $|w|^2+|x|^2+|y|^2+|z|^2=1$, and the
subscript $L$ stands for Lipkin.
The algebra, consists of considerations of the $U$-spin, $V$-spin and the
isospin as well as the orthogonality of states, is straightforward but
cumbersome and will not be repeated here.
The result is,
\begin{equation}
(|w|^2,|x|^2,|y|^2,|z|^2)=(\textstyle{3\over8},\textstyle{1\over24},
\textstyle{1\over48},\textstyle{9\over16}).
\end{equation}
As expected, the Lipkin states is mainly a $\Lambda$ bound state while the
$\Sigma$ contributions are small ($|x|^2,|y|^2\ll|z|^2$).
By isospin symmetry, the same conclusion holds for $|\bar Qsudd\rangle$.
We can also do the same decomposition for the $S=2$ $I=0$ state
$|\bar Qssud\rangle_L$.
\begin{equation}
|\bar Qssud\rangle_L = w_s|\bar Qu\rangle |\Xi^-\rangle
+x_s|\bar Qd\rangle |\Xi^0\rangle + y_s |\bar Qs\rangle |\Sigma^0\rangle
+z_s|\bar Qs\rangle |\Lambda\rangle.
\end{equation}
The results are
\begin{equation}
(|w_s|^2,|x_s|^2,|y_s|^2,|z_s|^2)=(\textstyle{1\over8},\textstyle{1\over8},
0,\textstyle{3\over4}).
\end{equation}
This state is again predominantly a $\Lambda$ bound state.
As a result, these three Lipkin states are expected to be stabilized by the
``Gell-Mann--Okubo mechanism'', which makes them lighter than other pentaquark
states with the same strangeness by (a fraction of)
$M_\Sigma-M_\Lambda=78$~MeV.
For a weakly bounded system, this can be a huge increase in stability.
Thus these Lipkin states are the ones most likely to survive the
destabilizing perturbations, which appear in higher orders in the bound
state picture.

Many properties of the Lipkin states can be predicted in the bound state
picture.
They consist of a $S=1$ isodoublet and a $S=2$ isosinglet, which look like
an SU(3) triplet (or, if the $S=0$ pentaquarks also exist, an SU(3) sextet)
but is in fact part of a ${\bf 15}$.
With $s_\ell=0$ and same parity as the normal heavy baryon bound state,
they have $J^P={1\over2}^+$.
The binding energy ${1\over3}V_0$ can be extracted from the heavy baryon
sector.
In the charm sector, we have this estimate of $V_0$.
\begin{equation}
\textstyle{1\over4}(M_D+3M_D^*)+M_N-M_{\Lambda_c}
=(1973+938-2285) \hbox{ MeV}=626 \hbox{ MeV},
\end{equation}
which gives ${1\over3}V_0=-209$~MeV.

The masses of the pentaquark states are given by the hamiltonian,
\begin{equation}
H_L=H_{heavy}+H_{baryon}+\textstyle{1\over3}V_0,
\end{equation}
where the heavy meson mass term $H_{heavy}$ should be taken as the
spin-averaged mass of the ground state pseudoscalar and vector mesons.
Note that, in our notation, $V_0$ is negative and hence the last term is
stabilizing.
Immediately this provides an estimation of the mass of the $S=0$ charmed
pentaquarks $|\bar cqqqq\rangle$.
\begin{equation}
|\bar cqqqq\rangle \sim 2702 \hbox{ MeV}.
\end{equation}
These states may have $(I,s_\ell)$ either $(1,0)$ or $(0,1)$.
\begin{mathletters}
The masses of the Lipkin states $|\bar csuud\rangle_L$,
$|\bar csudd\rangle_L$ and $|\bar cssud\rangle_L$ are also predicted,
\begin{equation}
|\bar csuud\rangle_L=|\bar csudd\rangle_L \sim 2857 \hbox{ MeV},
\end{equation}
\begin{equation}
|\bar cssud\rangle_L \sim 3009 \hbox{ MeV}.
\end{equation}
\label{cm}
\end{mathletters}
For comparison the masses of the $|\bar csuud\rangle$ state in the ${\bf 3}$,
${\bf \bar 6}$ and the $I={3\over2}$ state in the ${\bf 15}$ have also been
calculated.
\begin{mathletters}
\begin{equation}
|\bar csuud\rangle_3\sim 2896 \hbox{ MeV},\quad
(|w|^2,|x|^2,|y|^2,|z|^2)=(\textstyle{3\over8},\textstyle{3\over8},
\textstyle{3\over16},\textstyle{1\over16}),
\end{equation}
\begin{equation}
|\bar csuud\rangle_{\bar 6}\sim 2890 \hbox{ MeV},\quad
(|w|^2,|x|^2,|y|^2,|z|^2)=(\textstyle{1\over4},\textstyle{1\over4},
\textstyle{1\over8},\textstyle{3\over8}),
\end{equation}
\begin{equation}
|\bar csuud\rangle_{15\;I={3\over2}}\sim 2957 \hbox{ MeV},\quad
(|w|^2,|x|^2,|y|^2,|z|^2)=(0,\textstyle{1\over3},
\textstyle{2\over3},0).
\end{equation}
\end{mathletters}
Indeed the ``Gell-Mann--Okubo mechanism'' is at work: the Lipkin states are
pushed down below the ${\bf 3}$ and the ${\bf \bar 6}$ while the $I=
{3\over2}$ state in the ${\bf 15}$ is pushed up.
The analogous comparisons for the $|\bar cssud\rangle$ states are,
\begin{mathletters}
\begin{equation}
|\bar cssud\rangle_3\sim 3060 \hbox{ MeV},\quad
(|w_s|^2,|x_s|^2,|y_s|^2,|z_s|^2)=(\textstyle{3\over8},\textstyle{3\over8},
0,\textstyle{1\over4}),
\end{equation}
\begin{equation}
|\bar cssud\rangle_{\bar 6}\sim 3079 \hbox{ MeV},\quad
(|w_s|^2,|x_s|^2,|y_s|^2,|z_s|^2)=(\textstyle{3\over8},\textstyle{3\over8},
\textstyle{1\over4},0),
\end{equation}
\begin{equation}
|\bar cssud\rangle_{15\;I=1}\sim 3066 \hbox{ MeV},\quad
(|w_s|^2,|x_s|^2,|y_s|^2,|z_s|^2)=(\textstyle{1\over8},\textstyle{1\over8},
0,\textstyle{1\over4}),
\end{equation}
\end{mathletters}
Again the Lipkin state merges as the lightest one.

The analysis in the bottom sector is similar.
The estimate of $V_0$ from the $\Lambda_b$ mass
\begin{equation}
\textstyle{1\over4}(M_B+3M_{B^*})+M_N-M_{\Lambda_b}
=(5314+938-5641) \hbox{ MeV}=611 \hbox{ MeV},
\end{equation}
agrees nicely with that in the charmed sector.
\begin{mathletters}
With ${1\over3}V_0=-204$~MeV we obtained these predictions,
\begin{equation}
|\bar bsuud\rangle_L = |\bar bsudd\rangle_L \sim 6203 \hbox{ MeV},
\end{equation}
\begin{equation}
|\bar bssud\rangle_L \sim 6355 \hbox{ MeV}.
\end{equation}
\label{bm}
\end{mathletters}

Eqs.~(\ref{cm}) and (\ref{bm}) are the main results of this paper.
The bound state picture agrees with Lipkin's model that heavy pentaquarks
probably exist.
It should be noted that Lipkin's model is dictated by the principle of
``maximal flavor antisymmetry'', in which the most stable states will have
the wave function in the flavor space maximally antisymmetrized.
In the baryon octet, the only state satisfying this criterion is the
$\Lambda$.
In this light, it is not surprising that our results that the stable states
as heavy meson bounded to predominantly $\Lambda$ baryons agree so well with
his.

It must be stressed that, even within our model, we have {\it not\/} proven
the non-existence the non-Lipkin pentaquark states.
For example, the states in the ${\bf \bar 6}$ may as well have negative
binding energies.
These states, however, are expected to be more massive than the Lipkin
states and hence will decay to the Lipkin states electromagnetically.
Hence they are expected to be short-lived (when compared to
$\tau_{\rm Lipkin}\sim 10^{-12}$ sec) and much wider than the Lipkin states.

On the other hand, there are important corrections to our predictions which
have not been incorporated in our model.
For example, the hyperfine splittings $M_{D^*}-M_D=146$~MeV, which is an
$1/m_c$ correction, is significant for the stability of a pentaquark, the
binding energy of which is typically of the order of 100 MeV.
The effect of hyperfine splitting has been taken into account in our
calculations, but the effects of other $1/m_c$ corrections may also be
important.
Another potentially dangerous source of correction comes from the terms
in the chiral lagrangian with more than one derivatives.
It is not clear to the author how these effect can be estimated.

In conclusion, in the bound state picture we have calculated the spin,
isospin and parities of the charmed pentaquarks with $S=0$, 1 and 2 and
also estimated their masses.
All these predictions await experimental verifications, possibly by E791
at Fermilab.
Since these pentaquarks are stable with respect to QCD, they must decay
weakly.
If we assume the $\Lambda$ is just a spectator in the decay, a reasonable
assumption for a weakly bounded system, then the pentaquark life time will
be related to that of the constituent heavy meson.
\begin{mathletters}
\begin{equation}
\tau_{\bar Qsuud}\sim\tau_{\bar Qu},
\end{equation}
\begin{equation}
\tau_{\bar Qsudd}\sim\tau_{\bar Qd},
\end{equation}
\begin{equation}
\tau_{\bar Qssud}\sim\tau_{\bar Qs}.
\end{equation}
\end{mathletters}
In particular, the relation $\tau_{D^0}<\tau_{D^s}<\tau_{D_+}$ is expected
to translate into $\tau_{\bar csuud}<\tau_{\bar cssud}<\tau_{\bar csudd}$
in the pentaquark sector, as the valence $s$ and $u$ quarks can lead to
$\bar cs$ annihilation diagrams and $W$ exchange diagrams, which interferes
constructively with the spectator diagrams.
Decay channels like $(\Lambda D)\to\Lambda K\pi$ can be measured by
identifying the decay products, and the charmed pentaquarks may appear as a
peak in the $\Lambda K\pi$ invariant mass plot.

\acknowledgments
I must thank S. Kwan for urging me to generalize the results in Ref.~\cite{10}
to include pentaquarks with non-zero strangeness, and T.M. Yan for many
helpful discussions.


\begin{references}
\bibitem{1} R.L. Jaffe, Phys. Rev. {\bf D15} 267 (1977).
\bibitem{2} R.L. Jaffe, Phys. Rev. {\bf D15} 281 (1977).
\bibitem{3} R.L. Jaffe, Phys. Rev. Lett. {\bf 38} 195 (1977).
\bibitem{4} C. Gignoux, B. Silvestre--Brac and J.M. Richard, Phys. Lett.
{\bf B193} 323 (1987).
\bibitem{5} H.J. Lipkin, Phys. Lett. {\bf B195} 484 (1987).
\bibitem{6} S. Fleck, C. Gignoux, J.M. Richard and B. Silvestre--Brac,
Phys. Lett. {\bf B220} 616 (1989).
\bibitem{7} S. Zouzou and J.M. Richard, Few Body Syst. {\bf 16} 1 (1994).
\bibitem{8} Y. Oh, B.Y. Park and D.P. Min, Phys. Lett. {\bf B331} 362
(1994).
\bibitem{9} M.A. Moinester, D. Ashery, L.G. Landsberg and H.J. Lipkin,
Proceeding of Charm 2000, the Future of High Sensitivity Charm experiments,
275 (1994).
\bibitem{10} C.K. Chow, Phys. rev. {\bf D51} 6327 (1995).
\bibitem{11} G.S. Adkins, C.R. Nappi and E. Witten, Nucl. Phys. {\bf B228}
552 (1983).
\bibitem{12} M.B. Wise, Phys. Rev. {\bf D45} 2118 (1992).
\bibitem{13} T.M. Yan, H.Y. Cheng, C.Y. Cheung, G.L. Lin, Y.C. Lin and
 H.L. Yu, Phys. Rev. {\bf D46} 1148 (1992).
\bibitem{14} G. Burdman and J.F. Donoghue, Phys. Lett. {\bf B280} 287 (1992).
\bibitem{15} Z. Guralnik, M. Luke and A.V. Manohar, Nucl. Phys. {\bf B390}
474 (1993).
\bibitem{16} E. Jenkins, A.V. Manohar and M.B. Wise, Nucl. Phys. {\bf B396}
27 (1993).
\bibitem{17} E. Jenkins, A.V. Manohar and M.B. Wise, Nucl. Phys. {\bf B396}
38 (1993).
\bibitem{18} M. Gell-Mann, CTSL-20 (1961) unpublished.
\bibitem{19} S. Okubo, Prog. Theor. Phys. {\bf 27} 949 (1962).
\end{references}
\end{document}